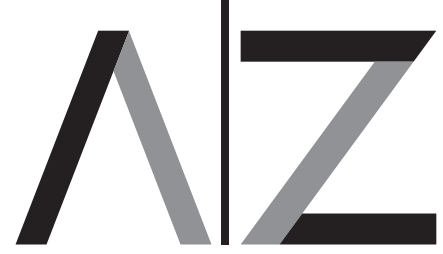

# Development and applications of Generative AI in architectural design studios

**Leman Figen GÜL[1], Burak DELİKANLI[2]*, Oğulcan ÜNEŞİ[3], Ertuğrul Ömer GÜL[4]**

[1] fgul@itu.edu.tr • Department of Architecture, Faculty of Architecture, Istanbul Technical University, Istanbul, Türkiye
*ORCID: 0000-0001-9374-4620*
[2] burak.delikanli@itu.edu.tr • Department of Architecture, Faculty of Architecture, Istanbul Technical University, Istanbul, Türkiye
*ORCID: 0000-0002-5576-306X*
[3] unesi23@itu.edu.tr • Architectural Design Computing Program, Graduate School, Istanbul Technical University, Istanbul, Türkiye
*ORCID: 0000-0002-4297-5194*
[4] e.gul@unsw.edu.au • School of Computer Science, Faculty of Engineering, University of New South Wales, Sydney, Australia
*ORCID: 0009-0009-2106-4357*

** Corresponding author*

**Abstract**
Recent advancements in deep generative models, with their capacity to yield outputs that are both visually appealing and semantically coherent, have served to further intricate the nexus between Artificial Intelligence (AI) and creativity. This integration of these models into creative disciplines, most notably within the domain of design, has been rapidly escalating. Nevertheless, a prevailing gap remains in our understanding of the impact of Generative Artificial Intelligence (GenAI) on design creativity. To address this gap, a longitudinal study was initiated, encompassing the development of a user-friendly GenAI interface. This interface was designed to facilitate the integration of GenAI models into design education, following a cyclical process of experimentation and refinement. The study involved the utilization of GenAI models in design studios, the collection of feedback, the development and evaluation of the GAI-A platform, and subsequent implementation in design education. The findings of the study provide a discussion based on the views and observations of students on the extent to which they use GenAI models, their perspectives and expectations of these models, and the extent to which this process improves their creativity in the context of the final product.

## 1. Introduction

In the 2004 movie I, Robot inspired by Isaac Asimov's book of the same name, the main character, Detective Spooner, asks the robot Sonny: "Can a robot write a symphony? Can a robot turn a canvas into a beautiful masterpiece?" Sonny, with genuine curiosity, responds "Can you?" This thought-provoking exchange underscores the rarity of creativity, even among humans (Gül et. al. 2024b).

Since Alan Turing's (1950) seminal work 'Computing Machinery and Intelligence', the field of artificial intelligence (AI)—which seeks to understand the mechanisms behind human intelligence and construct meaningful frameworks—has evolved through various eras shaped by technological advancements and theoretical debates.

Over the past seven decades, this journey has led to numerous groundbreaking—yet often controversial—innovations in the field. In the last decade, emerging AI methodologies, including advanced deep generative models such as Variational Auto-Encoders (VAEs) (Kingma & Welling, 2013), Generative Adversarial Networks (GANs) (Goodfellow et al., 2014), and Diffusion-based Models (Sohl-Dickstein et al., 2015), have shown immense potential to transform creative industries such as art, design, and architecture. In addition, the proliferation of advanced technologies and software has positioned itself as an invaluable aid, significantly augmenting the realms of creative industries. These technological advances not only facilitate but also enhance the creative process, establishing themselves as indispensable tools for practitioners in these fields.

Considering the recent technological impact, the rise of Generative Artificial Intelligence (GenAI) is poised to transform architectural design practices (As & Basu, 2021). Yet, the full extent of this impact remains unclear, as it appears to surpass even the fastest technological advancements (Hu et al., 2021). Initially striving to emulate human authenticity with compelling arguments, AI has now progressed to a stage where it can synthesize meaningful patterns from large datasets (Naveed et al., 2023). Besides, it finds extensive application in platforms that consistently generate content, such as blogs and social media, and is recognized as a time-saving tool for routine tasks (Davenport & Mittal, 2022). In these applications, GenAI has the capacity to function either as a valuable collaborator or a potential challenge to originality. Similarly, discussions on whether AI possesses creativity or consciousness are pertinent to this context (Leach, 2022; Boden, 1998).

Furthermore, the significance of GenAI models has increased, particularly in automating text and visual content creation processes. At present, 77% of cutting-edge tools incorporate some form of AI, and projections suggest that AI's global market share will reach approximately $60 billion by 2025 (Wardini, 2024). Consequently, the adoption of GenAI in the visual arts and creative industries is on the rise (Yakar, 2020).

In this context, exploring the potential applications of Big Data and AI technologies to facilitate, integrate, and enhance visual creative work in art and design communities is considered a valuable endeavor (She, 2019). The proliferation of GenAI models, particularly in creative industries like architectural design and design education, inevitably impacts these fields (Liao et al., 2024). In the context of education, similar to other technological developments, researchers pointed out that AI would have a transformative impact on education in many ways, through "transform[ing] processes, practices and institutions of teaching and learning" (Williamson et al., 2023, p. 1). Questions surrounding the utilization and management of these models, their impact on routine tasks such as data gathering and compilation, their potential role in decision-making, and the ethical and authenticity concerns they evoke, are currently on the agenda.

This study primarily focuses on investigating integration scenarios for GenAI models, capable of generating text-to-text, text-to-img and img-to-img, within the architectural design education. This study centers on students' perceptions of their own cre-ative actions and outputs within design, particularly how they assess these in relation to their design processes involving GenAI models, rather than defining creativity itself. In addition, the research delves into analyzing design students' approaches and usage of GenAI models, emphasizing both the opportunities and threats posed by these models. Over three academic semesters, with several case studies, various AI-assisted experiences were implemented during design studios with students, and the outcomes of our initial study were used to propose a user-friendly interface (named GAI-A), discuss the roles in design education, limitations, context, and potential challenges associated with GenAI models.

## 2. Background

### 2.1. Creativity in design education context

Creativity is generally understood as a capacity to produce "useful, satisfying" outcomes (Stein, 1953), "effective, innovative ideas" (Amabile & Pratt, 2016; Amabile, 2018), and products. Scientific disciplines differ on how creativity can be achieved, recognized, or appreciated (Gaut, 2014). The concept of creativity has a rich background, especially in psychology, with over 60 definitions based on various artistic perspectives (Sternberg, 2008). Cropley (2015) relates creativity to the concept of utility, arguing that an uncreative idea lacks utility for others. Additionally, numerous researchers have tried to define creativity using diverse criteria, including "high quality" (Sternberg & Lubart, 1998), "deviation from standards" (Niu & Sternberg, 2002), "non-commonality" (Simonton, 2012), and "aesthetics and authenticity" (Kharkhurin, 2014). While compelling arguments have been made for educating for creativity and in a way that can be described as creative (Gaut, 2014), past research has noted that explicit instruction in creative processes is more evident in disciplines such as the arts and humanities (Daly et al., 2016) and design (Gero, 2020). Within the design profession, the issue of creativity has long been a focus of research interest (see Cascini et al. 2022; Akın & Akin, 1996; Dorst & Cross, 2001; Gero, 1996; 2020; Shneiderman et al., 2006; Thompson & Lordan, 1999). For a design point of view, a creative design has to be novel, original or new and useful, or functional or valuable (Christiaans, 2002), as pointed out in definitions of creativity (Weisberg, et al. 2021).

Furthermore, creativity has a different meaning depending on the associated design profession (Diakidoy & Kanari, 1999; Dejean et al., 2018). For instance, creativity for an industrial designer may be a new technology, a new use, or a new material, while in another design field, an innovative approach may be considered creative. Koehler and Mishra (2008) state that a creative idea or product demonstrates aesthetic sensitivity and introduce the criterion of "integrity." Besides, Boden (2004) and Stein (1953, pp. 311–312) emphasize that evaluation of creative work based on its context, time, and location, considering whether it challenges conventional norms. Bryant and Throsby (2006, p. 508) consider it as the ability of individuals to generate ideas and products beyond conventional problem-solving methods using various thinking processes. Consequently, creativity emerges as a human capacity influenced by various factors in diverse contexts and creative work inherently demands novelty and originality.

Throughout history, creativity has been seen as a unique and mystical talent, and today, it is considered a teachable and learnable skill (de Bono, 2015; Barak & Goffer, 2002; Karni & Shalev, 2004). Numerous scholars posit that it is incumbent upon educational institutions to cultivate the creative capacities of their students (Marquis & Vajoczki, 2012). Additionally, a myriad of these institutions recognize creativity as a fundamental attribute in their graduates (Osmani et al., 2015). Research indicates that the incorporation of creativity in design education tends to be indirect rather than direct (Christiaans & Venselaar, 2005; Oxman, 2004; Rodgers & Jones, 2017). While such claims are prevalent, there is insufficient empirical evidence supporting the assumption that design education is effective in helping students develop creative design cognition (Gero et al, 2019). Additionally, there's limited research on design students' personal encounters with creativity in their education, highlighting a knowledge gap in how to effectively cultivate creativity in design education (Mclnerney, 2022). In this context, the present research endeavors to contribute to the extant literature by understanding AI-based design processes and deciphering their potential, based on students' perception of the creativity of their own designs.

**2.2. Generative AI, creativity, and architectural design**

A significant part of the discourse surrounding AI has centered on the concept of creativity, exhibiting a notable surge in attention (Du Sautoy, 2019; Miller, 2020). Boden (2009) defines creativity in the context of AI as "the ability to generate novel and valuable ideas" and provides an explanation of three ways for the creation of new ideas in relation to human creativity (Boden, 1998, p.348). The first of these is "combinational" creativity, which involves the generation of novel combinations of familiar ideas, including poetic imagery and analogies that are developed or explored. Secondly, there is "exploratory" creativity, which involves delving into the overlooked potential of conceptual spaces, resulting in structures ('ideas') that, whilst not necessarily novel, are unexpected. The thirdly, there is "transformational" creativity, which involves orchestrating changes that unlock the generation of ideas by transforming one or more dimensions of space, thereby generating new structures that were previously deemed impossible (Boden, 1998). It is posited that distinct amalgamations of the aforementioned three methods, which have been associated with human creativity, can also form an approach to the creativity of computational systems. The present article aims to explore how the application of GenAI in architecture studios can assist students in the design process, based on these approaches to the concept of creativity.

Following a decade of observation of the outcomes of deep generative models (Shi, 2022), there is an increasing inclination to embrace the notion of computer-generated creativity. This tendency is dramatically different from the skepticism that characterized the early days of computational journey (Leach, 2022).

The intersection of AI and design has been a subject of keen interest among researchers since the 1980s. The comprehensive literature review of Zwierzycki (2020) on AI research in architectural design reveals that studies are clustered around the topics of knowledge representation, design automation and layout generation. According to Zwierzycki (2020), none of the studies examined included "creativity" as a keyword. Only a few studies showed that architectural design is a highly relevant domain for the study of creativity (see Chan, 2015; Goldschmidt, et al. 2016) and yet, there is limited number of studies investigating students' training in the architectural design studio context with regard to creativity (see the review Casakin and Wodehouse, 2021).

One factor contributing to this oversight could be the gap between AI-generated products and those traditionally considered creative. The advent of GenAI models, exemplified by end-user applications such as ChatGPT, DALL-E (OpenAI, 2022; 2021), Midjourney (2022), and Stable Diffusion (StabilityAI, 2022) has precipitated a paradigm shift in the realm of AI. These models have garnered significant popularity and acclaim, prompting a surge in their deployment across diverse domains. The advent of these models has led to a surge of interest in investigating their potential to enhance creative output and facilitate their integration into design processes (Ploennigs & Berger, 2023a; 2023b).

Additionally, this paradigm shift highlights the growing recognition of the potential synergy between AI and creativity in design contexts (Ploennigs & Berger, 2023a). Research into GenAI has explored associations with various aspects, including integration into architectural design (Danchenko, 2020), application in conceptual design (Castro et al., 2021), exploration of architectural forms (Eroğlu & Gül, 2022), development and realization of sketching (Zhang et al., 2023; Tong, Türel et al., 2023; Yang et al., 2023), design ideation (Tholander & Jonsson, 2023), identification of design goals (Çalışkan, 2023), generating layouts and plans (Uzun et al., 2020; Ploennigs & Berger, 2023c; Aalaei et al., 2023), associating with design education (Başarır, 2022; Dortheimer et al., 2023; Tong, Ülken et al., 2023; Çiçek et al., 2023), semantic representation of architecture and engaging large language models into the AEC industry (Rane et al. 2023). The objective of the present research is to address a significant gap in the existing literature by conducting a comprehensive evaluation of the impact of GenAI models on the design process. The investigation will be particularly focused on the creative endeavors of design students.

## 3. Framework of the research

This article presents an ethnographic study of employment of GenAI in design education complemented with semi-structured interview and survey data. The integration of ethnographic methodologies into architectural studies is heralded as a promising avenue for initiating novel investigative and design-oriented questions (Yaneva, 2018). Our research, inspired by Schön's examination of the educational practice (1987), seeks to ethnographically reveal 'thinking in action', thereby challenging the established, methodical, and linear approaches to knowledge that dominate professional academia. Echoing Cuff's (1992) work on understanding the realm of professional architects, the significance of ethnography is emphasized. Cuff argues for a deep understanding of current practices to inform sound improvements, stating, "If we are to offer sound advice about how architectural practice ought to function, we must first know more about how it functions now" (1992, p. 6). Similarly, to refine GenAI tools that enhance the creativity of design students, we must first understand their current working methods with them.

This study focused on the query of whether GenAI models can serve as beneficial tools for fostering creativity in architectural design education. The research explores the following questions:

- How do design students integrate GenAI models into their design process and at which stages do they interact with them? How can GenAI models support various stages of the design process?
- What perceived benefits do students associate with GenAI models, and to what extent do they believe these tools enhance their creativity?

In the present study, the theoretical framework underpinning design studios was integrated with Wrigley and Straker's (2015) five-stage Educational Design Ladder (EDL) model and GenAI models, with the adaptation of this framework occurring in alignment with the specific themes that characterized each studio. These skill development-focused processes were methodically adapted to align with the fundamental subject and theme of the respective applied studios. Due to the inherently nonlinear nature of the design process, the analysis-synthesis-evaluation procedures were executed in a cyclical manner. The following Table 1 illustrates the mapping of this theoretical framework and the research outputs.

### 3.1. Procedure

The theoretical underpinnings of our investigation trace back to the summer of 2022, characterized by the widespread integration of models such as DALL-E, ChatGPT (OpenAI, 2021; 2022), and Midjourney (2022). In this respect, the present research encompasses a series of longitudinal studies. These longitudinal studies commence with a process of understanding the current situation and expectations. To ascertain the potential of GenAI in architectural design studios, a series of design workshops spanning five semesters was conducted, we present the first three in this article. During the initial two semesters, students were encouraged to utilize existing GenAI tools within their design studio endeavors (in Section 3.1.1 and 3.1.2, e.g. Case I-II). Subsequently, drawing insights from our observations and discussions with students, we proceeded to develop an interface Generative Artificial Intelligence in Architecture (GAI-A). Throughout the third semester, emphasis was placed on integrating the developed GAI-A into students' design processes. This involved conducting sessions aimed at equipping students with the requisite skills and knowledge to effectively utilize the interface within their architectural design projects. These sessions served the dual purpose of data collection and iterative development of the GenAI interface, informed by students' feedback (in Section 3.1.3, e.g. Case III). In the following two semesters, an investigative endeavor was undertaken with a select cohort of students centered around the theme of design processes. This investigation is still in progress, and its findings shall be elaborated upon in a separate publication at a later stage. The present article herein delineates the cases that were handled in the initial three semesters of this study (Figure 1).

*Table 1. Conceptual framework of the design studios.*

| Adaptation of the Educational Design Ladder (EDL) model in architectural design studio with GenAI models (based on Wrigley and Straker, 2017) | | | | | |
|---|---|---|---|---|---|
| **EDL model phases** | **Learning Outcomes** | **Studio Themes** | **Environment /Medium** | **Outcomes** | **Research Output** |
| **1-Foundational Level: Knowledge comprehension** | Knowledge of basic concepts | Basic architectural design concepts | Seminars, site visits | Research and personal repository | Reflective journals |
| | | Basic concepts of GenAI | Hands on workshops | | Surveys |
| **2-Product Level: Application** | Comprehension of GenAI | Implementations, Executions and translations | Able to use GenAI Models and GAIA interface | Homeworks specific to problem definition and design problem space | Semi-structured surveys |
| | Design problem space explorations | | | | |
| **3-Analyse** | Knowledge of classifications, categorizations and generalizations | Incubation period, design problem space exploration | Sketching, collage and exploring with GenAI | Conceptual diagrams, mass design proposals, façade and style proposals, silhouette drawings | Observations, researchers notes |
| **4-Synthesis** | | Inspiration period, design solution space exploration | | | Independent jury evaluations |
| **5-Evalution** | Strategic knowledge on cognitive tasks and self-knowledge and self-reflection | Appraise, Value, Select | Desk critique, jury critique | Presentations | |

### 3.1.1. Initial phase: Case-I

This semester marked our initial exploration of the link between GenAI and design education. To facilitate this process, we invited students to delve into the potential of GenAI models through an introductory workshop (e.g. I, Figure 1) preceding the formal design brief, followed by monthly juries and weekly meetings throughout the semester. In our university, architectural design studios are regularly instituted, wherein project locations and themes are communicated to students through studio posters and course syllabi. The exploration of GenAI's impact on the architectural design process necessitates the identification of students willing to engage with these tools for an entire semester. This focus, communicated through both posters and syllabi, served to attract students manifesting interest, involving students at the third year, extended over a duration of 14 weeks, with a commitment of 8 hours per week.

### 3.1.2. Development phase: Case-II

In the second semester, to enrich the learning experience, we introduced a comprehensive workshop module (e.g. II, Figure 1), building on the insights gained from the previous semester. Over five sessions, students engaged in the design of future educational spaces, utilizing spatial computing (AR-VR) for design representation and GenAI models for design inspirations, including ChatGPT, DALL-E (OpenAI, 2022; 2021), and Midjourney (2022).

### 3.1.3. Test phase: Case-III

Throughout the third semester, students actively engaged with GenAI tools, employing text-to-text, text-to-img, img-to-text, and img-to-img functionalities in their design processes. Consecutive seminars and subsequent workshops (e.g. IV-VI,

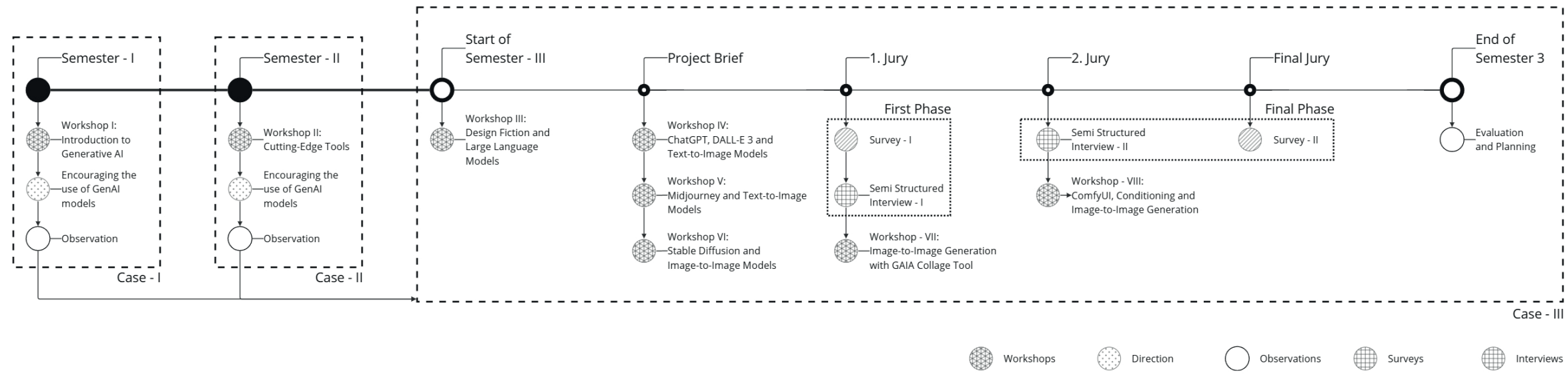

***Figure 1.*** *Research framework during the three academic semesters.*

Figure 1) formed an integral part of the program, featuring guest speakers experienced in using GenAI models. Essential concepts such as prompt engineering and negative prompts were elucidated, accompanied by practical guidance on utilizing end-user tools like DALL-E 3 (OpenAI, 2021), Midjourney (2022) and web applications such as LeonardoAI and OpenartAI which were based on existing models such as Stable Diffusion and DALL-E. Final seminar delved into more intricate examples and advanced rendering models, employing Google Colab, and emphasizing the img-to-img technique, with a majority based on the Stable Diffusion (Stability AI, 2022). Each seminar was followed by swift tutorials and hands-on sessions facilitated through a Miro (2019) board (Figure 2).

After half of the third semester, a pair of consecutive workshops (e.g. VII-VIII, Figure 1) were meticulously organized to test our developed GenAI interface (GAI-A). The first workshop delved into the intricacies of the "Collage Tool" within the GAI-A interface, while the second workshop centered around introducing an approach involving personalized workflows and delving into more advanced topics related to diffusion models. These sessions were strategically planned in response to ongoing observations throughout the semester and, more significantly, insights gathered from semi-structured interviews.

Students received foundational insights into Stable Diffusion, covering terminology, components, and parameters such as latent space, encoders, guidance scale, and schedulers. Additional information on custom workflows, fine-tuned models, LoRAs (Hu et al., 2021), ControlNets (Zhang & Agrawala, 2023), and Upscalers was provided to enhance students' proficiency in Stable Diffusion-based GenAI models, encompassing various parameters and capabilities. To apply this knowledge practically, students were introduced to ComfyUI (Comfyanonymous, 2023), a node-based interface. The familiarity of ComfyUI with Grasshopper (Grasshopper 3D, 2007) a widely recognized node-based visual scripting interface facilitated easier comprehension for students. ComfyUI's widespread use and exposed nature also proved valuable for students to access tutorials, user workflows, and community content outside of the seminar.

## 4. Developing the Generative Artificial Intelligence for Architecture (GAI-A) interface

### 4.1. Identifying students' needs

Conversations with students reveal an expectation for an easily navigable GenAI interface and a tool that streamlines the collage-making process, aligning with their individual styles and image preferences. They articulate that such a platform can enhance their design ideation, decision-making, and overall creative process. Thus, based on these findings from our work with students, we developed a GenAI interface for architectural design (GAI-A), which is easy to use for non-programmers. In subsequent stages, as additional components are incorporated, this platform is poised to become an ideal medium for individual designers engaged in architectural design and style studies.

### 4.2. Components of the GAI-A interface

The GAI-A system utilizes a convergence of sophisticated technologies, incorporating React.js for the frontend interface, Flask for backend operations, and a synthesis of img-to-prompt and prompt-to-img generation employing a diffusion model (Midjourney v5.2) and a Large Language Model (LLM) (ChatGPT-4). The core purpose of the system is to transform image collages into cohesive, singular representations that encapsulate the user's intended message or theme. We employ a fundamental client-server architecture, wherein the frontend communicates with a singular-instance web server application. These collages are subsequently transmitted to the server, a web application based on Flask, which handles the requests and engages with the requisite AI technologies.

Our system operates through a sequential pipeline of established technologies, employing both image-to-prompt and prompt-to-image creation (at Midjourney v5.2) alongside a LLM to generate an output image from an input collage (Figure 3). Users interact with the system through a simple Graphical User Interface (GUI), where they can create collages using uploaded images. Upon completing the collage, clicking the generate button prompts the user to wait as the image data is sent to the backend for processing. The backend processes the created collage and returns the output image to the GUI (Figure 3), allowing the user to observe and save it.

Initially, our process involves obtaining a textual description of the collage using image-to-prompt generation. This is achieved by utilizing the describe mode of Midjourney, generating four sets of prompts that describe the collage. These prompts are then fed into ChatGPT for selection and refinement. The role of the LLM is to select and refine a prompt suitable for Midjourney's imagine mode, ensuring the likelihood of producing a desirable final image. This step also ensures that the final image generated is not another collage but a conclusive 'render' representing the user's intended idea. Once we acquire a refined prompt, it is input into Midjourney in imagine mode, resulting in the production of our final image. The GAI-A applica-tion is distinguished by its capacity to consolidate a multitude of workflows and models within a streamlined interface, exhibiting a rapid learning curve. A notable feature of the application is its ability to seamlessly integrate model parameters into the system, a process that is informed by student feedback (see Gül. et. al. 2024a, for more information about the interface). Furthermore, we have leveraged this capability to incorporate specialized Stable Diffusion models and workflows in subsequent versions, enhancing the application's versatility and functionality.

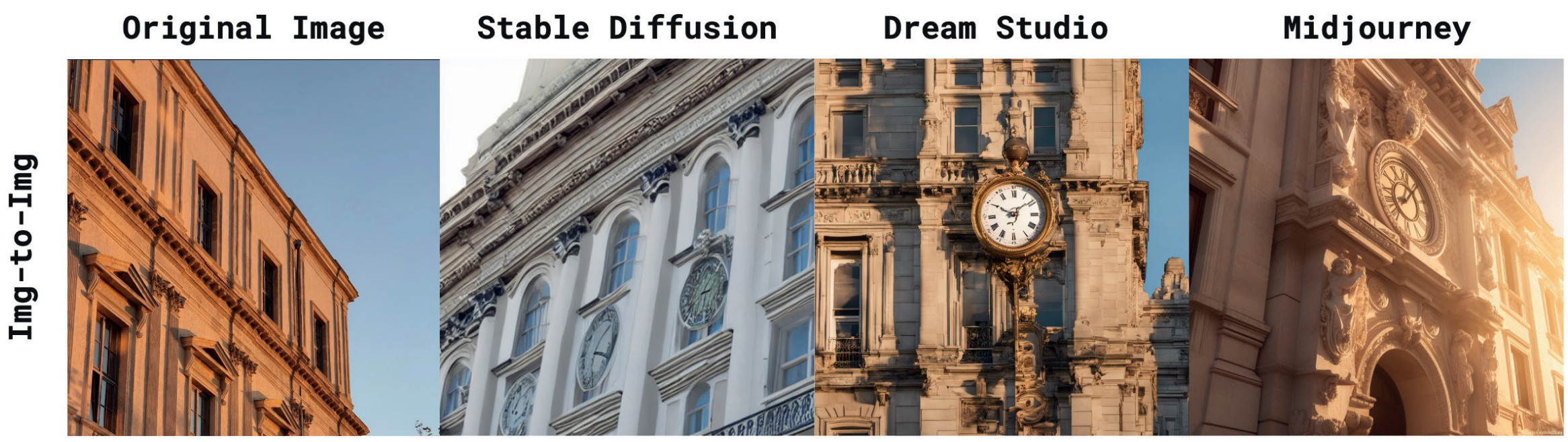

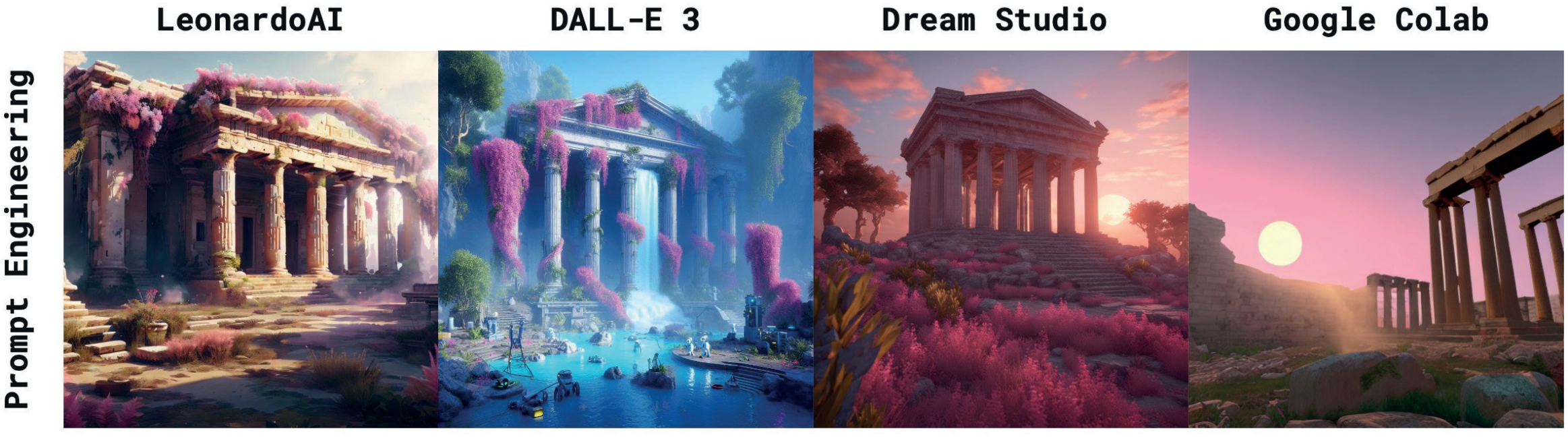

***Figure 2.*** *Examples of student-generated images and prompts in Workshop IV, V, and VI This collaborative board allowed students to share the images they generated along with the corresponding prompts.*

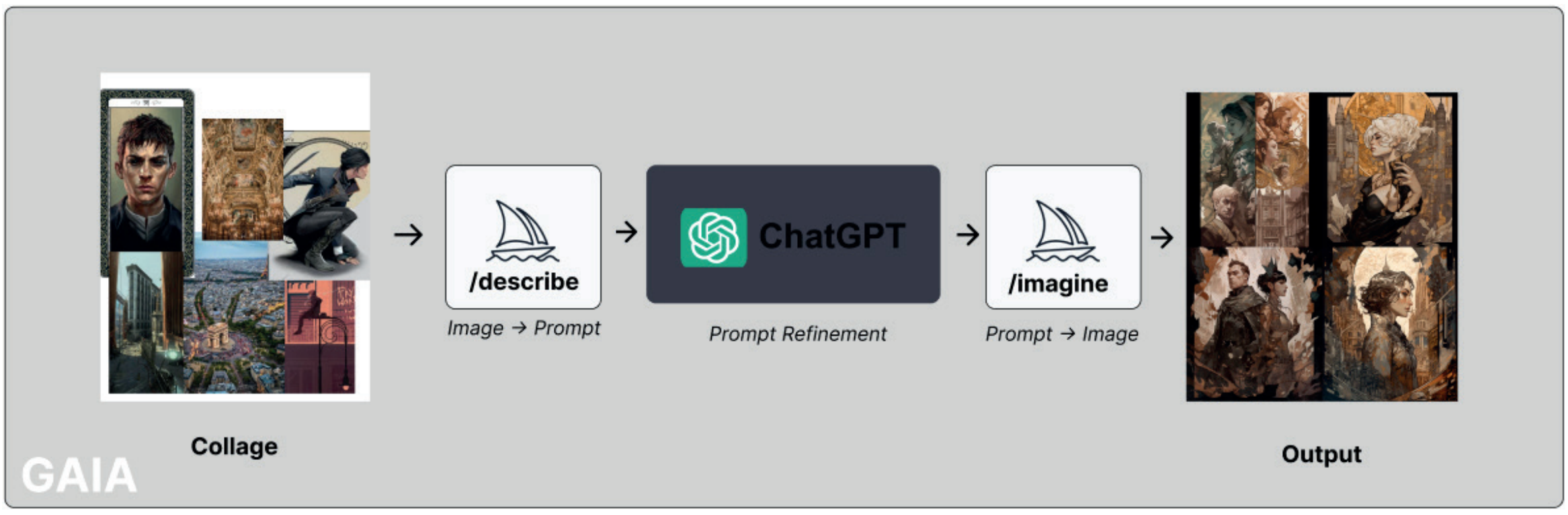

***Figure 3.*** *Input and output generation process of the GAI-A Collage.*

### 4.3. Data collection and coding

Data collection methods include observation, subjective reporting, evaluation of final products and processes, system logs, semi-structured surveys and questionnaires. We collected semi-structured surveys and questionnaires only in the third semester (Case-III), there were 24 students enrolled in the Course, all having 3 years of experience in architectural design. The call for participation in the study was extended to all students attending the workshops and was entirely voluntary. Prior to engaging in the surveys, which were conducted in two stages—the initial survey during sixth week (N=18) and the final survey during thirteenth week (N=24)—all participants provided informed consent. Furthermore, semi-structured interviews were undertaken with a cohort of students during each phase (N=6), aiming to illuminate individual aspects of their design processes and stages.

The questions of survey and semi-structured interview were formulated with the objective of discerning students' self-perceptions and evaluating the extent of their creative capacities using GenAI models in design. To understand how students assess their own architectural production and design processes and to what extent they think the GenAI models are used to support their creativity, a questionnaire was designed by adapting the question set developed by Diakidoy and Kanari (1999) (Table 2).

In the semi-structured surveys, we used Reflexive Thematic Analysis (Braun & Clarke, 2019) to learn about students' design processes, expectations from GenAI models and current usage trends. The questions, with the first one formally posed initially and the rest seamlessly woven into the conversation, were as follows:

- How would you describe your design process from start to finish; in an ordered, step-by-step manner?
- Which of the steps involving design are the ones that you produce the most and the least variations (or iterations for iterative tasks)?
- Which of the steps involving design are the ones that are the hardest and the easiest to produce variations (or iterations for iterative tasks)?
- Which of the steps in your design process are the hardest, most tedious or time-consuming? Are there any steps that you choose to skip because of the aforementioned qualities?
- Are you currently employing any AI tools or assistants in your project? If so, what, how and in which steps of your design process are you using them?
- Existing or not, what kind of generative artificial intelligence tools would you have preferred to use in your design process?

***Table 2.*** *Responses to the question of how do you define creativity (Diakidoy & Kanari, 1999).*

| ***Creativity is:*** |
|---|
| ***the ability to think in original ways*** |
| ***the ability to produce something unique, original*** |
| ***the ability to deal with situations and problems that have never been encountered before*** |
| ***the ability to create something original out of nothing*** |
| ***the ability to present something in a different way*** |
| ***the ability to influence the environment in an effective way*** |
| ***the ability to come up with new solutions and ideas in order to solve difficult problems*** |
| ***the ability to use whatever means and materials are available to make something that is useful in someway*** |
| ***the fulfilment of personal needs—as a way to express yourself*** |
| ***related to imagination—and that is why it can be developed given the right circumstances*** |
| ***a way of reacting to things that does not require too much thinking or knowledge fulfilling my potential*** |

The following categories were defined to be able to map the patterns. The correspondences of these categories in design process steps can vary in order. GenAI-related insights are noted inside each category (Table 3).

## 5. Results

### 5.1. Observations of design workshops

#### 5.1.1. Observations of case-I

In Semester-I students immersed themselves in the vibrant Cibali district of Istanbul, tackling the distinctive project titled "Designing the Black Box: Creating the Intimate Space of Arts". The studio's objective was to craft a comprehensive design for art events, emphasizing the intricate interplay of spaces and specialist functions, including public interface, assembly, and functional performance. This conceptual project, positioned at the nexus of creative industries and techno-cultural advancement, provided an ideal theoretical ground for experimenting with GenAI models. Notably, some students leveraged ChatGPT (OpenAI, 2022) to analyze their design objectives right at the onset of the creative process, extending its use to other coursework. A couple of students drew inspiration from DALL-E 2 (OpenAI, 2021) and Midjourney (2022) during the early stages of the design process. However, due to the developmental phase and the price of these tools, only one student seamlessly integrated GenAI throughout the entire design process (Figure 4). The notable aspect in this student's work lies in the innovative approach adopted during the design process. At a pivotal juncture, the student generated imagery by articulating ideas about the spaces, essentially making initial inferences before proceeding with the design of 3D spaces based on these conceptualizations. Figure 4 illustrates the progression of the relevant space through images captured from various locations. An insightful observation from this semester was the varying levels of student interest and familiarity with cutting-edge tools upon entering the studio. Recognizing this diversity, we underscored the importance of conducting workshops that not only capture student interest but also effectively introduce and familiarize them with these innovative tools.

#### 5.1.2. Observations of case-II

In Semester-II students studied in the educational complex of Germencik, Aydın, delving into the innovative project titled "Neo-Education: Designing the Future Education Spaces". The studio's objective was to create a schematic and developed design for education spaces, placing a strong emphasis on envisioning the future of learning environments, interrelationships of spaces, and specialist functions, including public interface, assembly, and functional performance, all within a cultural significance context. The proposal aimed to contribute to a knowledge-based educational context, nurture the quality of life, and speculate on new forms of active and value-added learning to support techno-cultural advancement.

Distinguishing itself from the previous semester, this studio focused on Neo-Education, integrating spatial computing (such as AR-VR tools), and providing an immersive experience with GenAI tools within fictional and virtual spaces. This forward-looking approach allowed their creativity to flourish, and the students' overall sentiment towards these tools was positive. A key factor in this positive reception was the perceived success of GenAI in constructing fictional spaces. Students expressed that these tools not only supported their imagination but also facilitated creation in the realm of the unknown.

A significant observation during this semester highlighted the importance of dynamic adjustments for students with different backgrounds in workshop modules. The level of integration of cutting-edge tools remained consistent when workshop outcomes were not continually measured or adapted to align with the evolving needs of the students throughout the semester. Thus, we emphasize the necessity of reorganizing workshop modules in response to the dynamic requirements of the students as the semester unfolds.

### 5.1.3. Observations of case-III

In Semester 3, students delved into the historical Kemeraltı, İzmir, taking on the distinctive project titled "Envisioning Future Urban Life: Infill Design in a Historical Urban Setting." This studio embarked on an exploration of the concept of the Design Fiction (Celik et al., 2023), with a primary focus on crafting future spaces within the context of urban heritage. The proposed designs aimed to establish community hotspots through the integration of cutting-edge alternatives. This approach, in turn, sought to catalyze socio-economic growth and add value through its programmatic elements.

***Table 3.*** *Coding categories of interviewees' design process expressions.*

| Code | Definition | Example of Use |
|---|---|---|
| **Site Analysis and Research** | Including field trips, analyses and information gathering through internet research. Common outputs are in the form of text, maps and diagrams. Mapping activities based on the site and the environment, and further identification of site-specific conditions are also included in this category. | GenAl models are not used in this category, however one participant stated that they wished an Al assistant could help them with their analyses. |
| **Defining Objectives** | Defining user scenarios, naming site-specific or studio-specific requirements and creating an architectural program. | LLM is commonly used in this step, assisting the participants in creating user scenarios, compiling analyzes and creating architectural programs. |
| **Design Research** | Researching architectural repositories, examining architectural styles, taking inspiration from various sources, and collecting information on materials and construction-related techniques. | Latent diffusion models are used to generate images to be taken inspiration from. They can also be used to help visualize specific conditions. |
| **Concept Design** | Including producing variations for the design proposal as well as activities such as sketching, initial positioning and rough mass modelling, expressing or gathering inspiration. | Interviewed participants did not show interest in using GenAl models in these steps. One participant with a physical model-making workflow stated that they might be interested in trying latent diffusion models for generating design concepts in order to relieve them of their physical workload. |
| **Design Development** | Commonly including iterative processes to improve the design over time. The design proposal is often volatile and susceptible to major changes as well as minor ones during these steps. | GenAl models are not used in this phase, although there is desire to use them. Participants expressed interest in using plan-generating and 3D-model-generating GenAl tools in the steps involved. |
| **Design Finalization** | Including a stage where proposals finalized and finer details are being brought in. Major changes are uncommon, minor ones can happen. These steps often include the production of the deliverables, or the materials to be presented in the form of 3D models and technical drawings. | GenAI tools are not used in this category but there is interest in using them. Participants explained that they would like to explore tools that could produce meaningful technical drawings and detailed 3D models in these steps. |
| **Presentation** | Including the post-processing and preparation of the final presentation. These steps often include rendering the design through various techniques, most often through producing rendered views of a 3D model, and post-processed plans, sections and elevations. The resulting materials are then often arranged in physical or digital posters. | Some of the final presentations, students shared initial renders form GenAl and GAIA as indicating of the design development phase. |

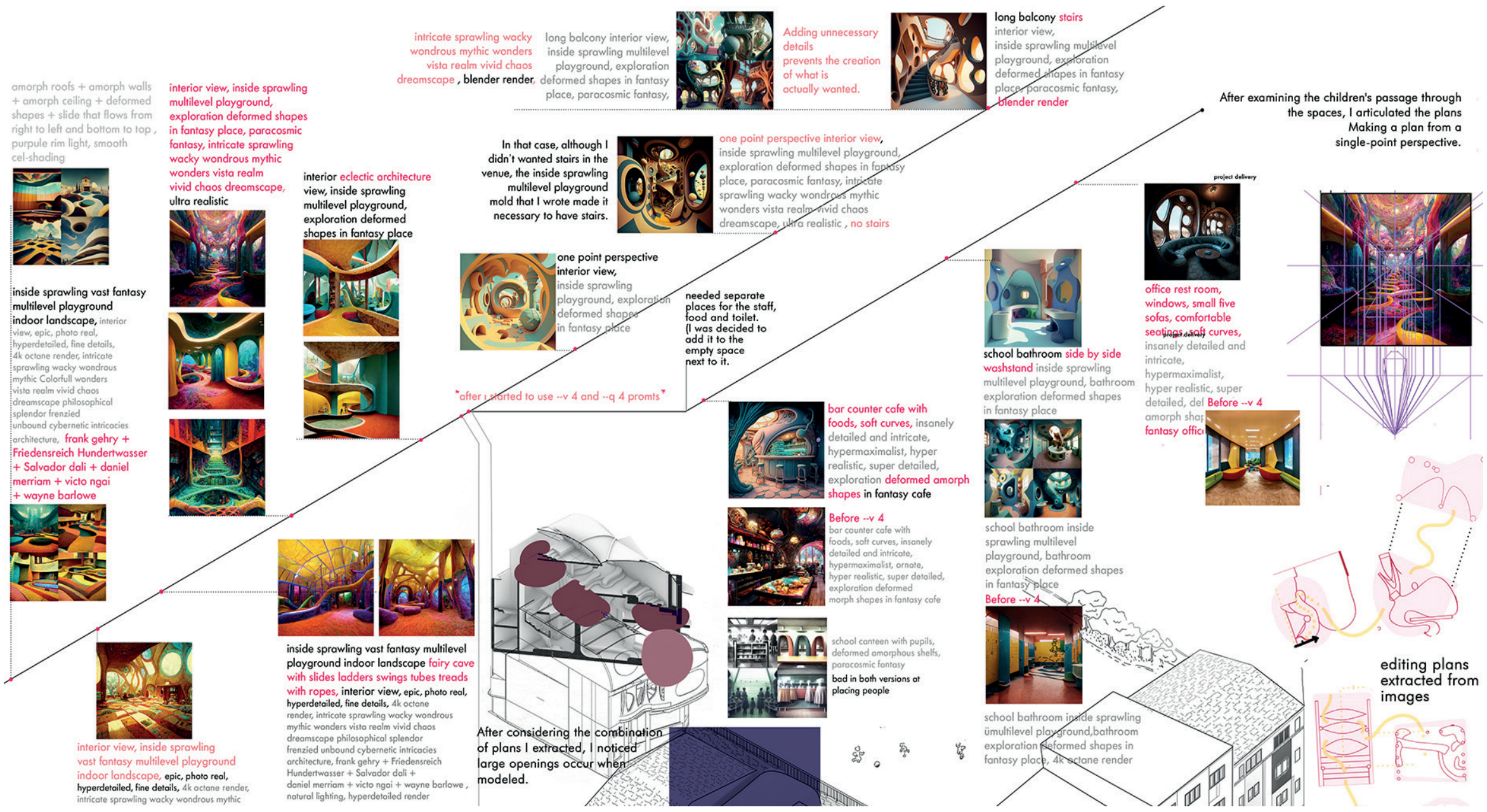

*Figure 4. An architectural design method integrated with Midjourney (Student work from Semester-I).*

Workshop VII provided students with the opportunity to create architectural collages with GenAI support, merging conceptual visuals from their projects (Figure 5). In Workshop VIII, a select group of students explored methods involving conditioning to generate results independently. This seminar and the subsequent workshop were tailored to address img-to-img generation and techniques involving ControlNets (Zhang & Agrawala, 2023) models. These methods were chosen for their capability to generate images with specific qualities, such as producing facades on mass models, restyling existing renders, or crafting stylized architectural sections and plans from raw inputs. Throughout this studio, students were provided the chance to actively engage with the outcomes, allowing them to experiment and fine-tune their prompts for optimal results. Notably, it became evident that incorporating generic architectural terms into the model did not consistently yield the desired outcomes. Instead, the emphasis shifted towards the necessity of articulating prompts with a specific stylistic preference, often involving references to distinct graphic artists or well-known architectural examples. Numerous workflows, aligned with the discussed topics, were demonstrated and distributed, followed by a hands-on session where results were uploaded to a collaborative Miro (Miro, 2019) board (Figure 5).

## 5.2. Results of surveys

The survey questions were categorized into two sections. The first section aimed to evaluate the assistance provided by the GenAI model used in the design process and its potential impact on the students' creative design process. In the second section, students were asked about the elements they believe should be incorporated into an interface designed for this purpose. Here we presented some of the results of the surveys, gathered using a 5-point Likert scale, as per below. Overall, the majority of students (95% of all students (N:18) in sixth week (e.g. Survey I, Figure 1) and 96% of all students (N:24) in thirteenth week (e.g. Survey II, Figure 1) feel that the used GenAI models positively affect their creativity during the design process, as shown in Figure 6. The surveys included the same questions designed to prompt self-assessment of students' thoughts and emotions while using AI, and in particular the used diffusion models in the design process.

Based on our discussions and observations with students, the prompts allow us to draw specific inferences related to the model employed. As evident from the students' outcomes, the images that align more closely with the prompts focused on the theme of the future, resembling stage-like creations seen in science fiction movie sets, yield more favorable results. In contrast, the prompts that reference architectural styles or local architectures don't yield the same desired outcomes.

This underscores the significance of the need of developing a model and / or an interface that empowers users to choose specifically trained models (checkpoints) according to needs in the context of architectural design education. This fundamental finding was very influential in our development of the GAI-A interface.

The feedback from users underscores the GAI-A interface's impact on fostering creativity in the design process. Many students appreciated its role in early design stages, where it effectively detected intricate details and contributed to the solidification of conceptual designs. During idea generation, they found it to surpass traditional methods, providing inspiration, aiding in visualization, and offering diverse perspectives.

Numerous students highlighted the tool's positive influence on their creative thinking processes, noting its ability to streamline work, generate different ideas, and expedite project development. The application's contribution to the program and analysis phase, as well as its role in encouraging broader and more detailed thinking during the idea stage, further emphasized its positive impact on creativity.

While some students enjoyed its application in the early design phases, others acknowledged its influence on motivation and creativity, emphasizing its positive effect on overall creative skills. Despite varying opinions on its rendering capabilities, they generally expressed optimism about the tool's future usability, foreseeing it as a valuable asset for faster production, continued creative development, and the generation of effective design solutions. The consensus was that the application's utility tends to grow with continued use, making it an increasingly essential tool for nurturing creativity in the design process.

### 5.3. Results of semi-structured interviews

The questions, designed as conversation starters, tactfully addressed aspects of the design process that are typically challenging, repetitive, and time-consuming. This approach aimed to prompt interviewees to consider how GenAI-based models could expedite these tasks and be integrated into their design workflows. Conducted during a studio session in the sixth week of the third semester (see Figure 1), the first phase of interviews involved participants (N=6) whose projects ranged from initial research to early concept sketches. Each interview lasted approximately 30 minutes and allowed participants to describe their design processes without being questioned about the completeness of their design. Although the small sample size and participants' limited experience preclude broad generalization, the insights provide a valuable foundation for examining emerging uses of GenAI in design.

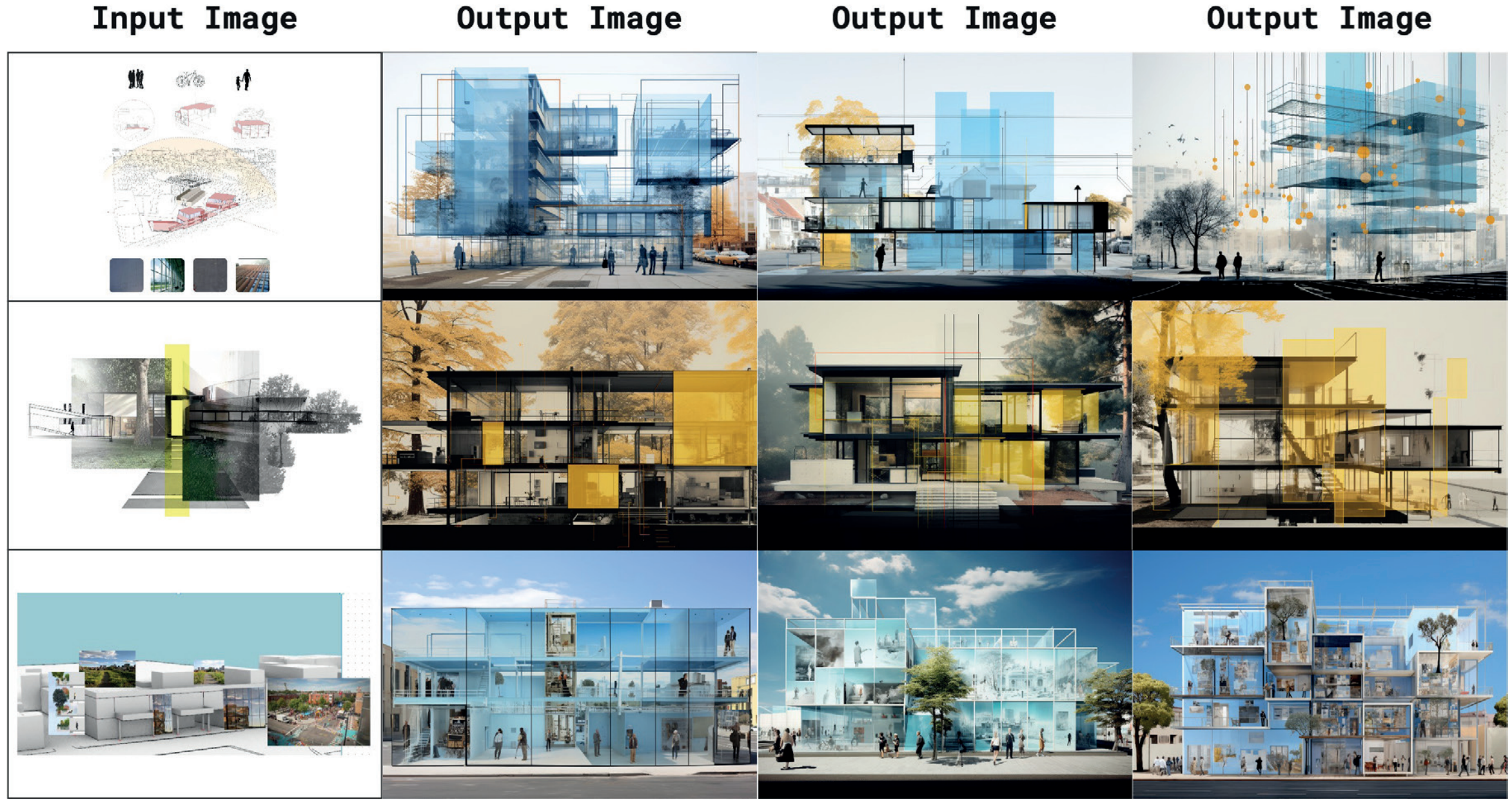

***Figure 5.*** *Examples of outputs, collage tool of GAI-A interface in Workshop VII.*

A thematic analysis of the first phase of interviews revealed several recurring steps in participants' design processes, despite varying terminology. These were organized for pattern exploration (see Table 3), clarifying shared approaches and illustrating potential GenAI-related strategies at each stage (Figure 7a). The final phase of interviews occurred in the thirteenth week (see Figure 1), once students were finalizing their designs and preparing final presentations (Figure 7b). By revisiting the same participants at the culmination of their semester-long projects, this follow-up captured changes in perceptions and practices related to GenAI. The additional experience some participants had gained with GenAI tools between the two interview rounds led to more in-depth and reflective discussion; students' active usage and interest of GenAI in the design process reflects preliminary insights (Figure 7a). However, after the workshops in the studio process, active GenAI usage is observed in more uncertain phases of the design (e.g. Problem Space, Figure 7b). In contrast, the active usage of GenAI either decreased or disappeared in the more certain phases of the design process (e.g. Solution Space, Figure 7b).

Across both interviews, participants consistently underscored the value of GenAI models in enhancing creativity, with many describing them as instrumental for visualizing abstract concepts and translating emerging ideas into concrete images. This was often characterized as "gathering inspiration" or "bringing ideas to life," particularly useful when dealing with unfamiliar elements or complex concepts. Nevertheless, participants remained cautious about relying on GenAI throughout the entire creative process. Their main concerns were the perceived imperfection or incompleteness of AI-generated outputs, prompting skepticism regarding the accuracy of some results and contributing to a reluctance to substitute AI for human judgment during the workshops.

Another frequently mentioned issue involved a perceived loss of ownership when employing GenAI. While participants benefited from rapid visualization and producing design materials beyond their own skill sets, some expressed hesitancy about depending solely on AI-generated assets. Instead, they preferred to modify or recreate them to preserve a personal connection with their work. This practice illustrates a balancing act between leveraging AI's capabilities and maintaining creative authorship. Despite these reservations, participants viewed GenAI tools as significant facilitators of design, enabling quicker exploration of multiple concepts and expanding their repertoire of potential solutions. Overall, these interviews suggest that GenAI can play a supportive yet bounded role in design processes, catalyzing ideation and efficiency while still requiring human reflection and adaptation to maintain both accuracy and personal creative identity.

### 5.4. Findings and conclusion of integrating GenAI into the design process

One notable finding is that design students are hesitant to relinquish control over the tools they use, a common sentiment echoed in our interviews. Proficiency in these tools correlated with the students' success in advancing their processes, a trend further underscored by the increasing frequency of workshops conducted throughout each semester.

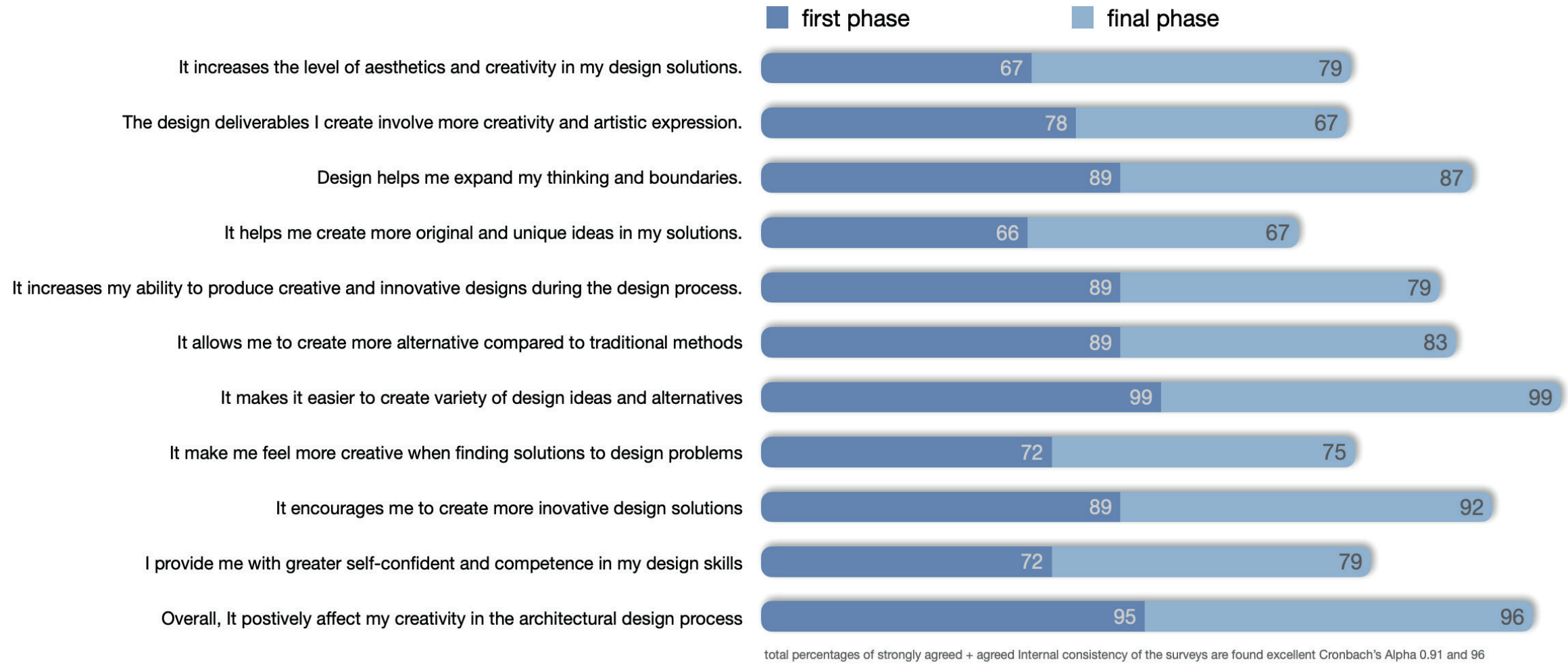

***Figure 6.*** *Students' self-perception of how used AI models supported their creativity in Architectural Design Studio in the first and final phases of their design process.*

It is, perhaps, unsurprising that the progression of the design process aligns with the mastery of the tools used, as highlighted by Robertson and Radcliffe (2009). In this context, we observed that the effectiveness of students' tool utilization was influenced by a careful scaffolding approach in teaching, as suggested by Ninio and Bruner (1978). Hence, it is plausible to assert that an augmented technical knowledge load (in terms of AI), facilitated by a collaborative process associated with the employed GenAI model, empowered the student to propose creative solutions more effortlessly during the design process. As Maher (2012) inquired, the question of whether the creative idea originates from the computer, a human, or emerges as a collaborative structure from the interaction between individuals and computational systems is particularly pertinent in our context.

In this paper, we sought to comprehend how architecture students employ GenAI to enhance their design processes and the manner in which they assess the efficacy of this collaboration in promoting their creativity. Boden's triad of creative processes has been observed to a certain degree in the students' processes. Specifically, the discovery process manifests in various forms. This discovery process, illustrated by the utilization of GenAI models for concept development and inspiration, particularly in the early design phase, is designated as 'exploring unknown problem space' through a range of stimuli. Smith (1995) underscores that the employment of diverse stimuli has the capacity to disrupt stereotypes, fostering association and innovation. In a similar vein, Goldschmidt and Sever (2011) posit that exposure to diverse stimuli contributes to the formation of concepts. In this particular instance, visual stimuli, particularly those generated by AI, hold potential as a useful tool. In exploring the design problem space, the students employed these tools to investigate possibilities in the solution space, to address the gaps in their conceptualization through LLM and GenAI, and to concurrently generate a source of inspiration. This approach was particularly evident in the development of ill-defined architectural programs, where the definition of ambiguous areas was facilitated. Visual stimuli were also theorized to stimulate subconscious creativity, thereby enhancing the creative thinking of designers (McKim, 1980). Goldschmidt and Smolkov (2004, 2006) state that designers have different preferences for sources of inspiration and that different types of visual stimuli can stimulate designer creativity in various ways. Given that appropriate visual stimuli have been demonstrated to significantly improve idea generation (Gilhooly, 2007), there is also a case for exploring the potential of visuals produced by GenAI to support the exploration process in the formation of creative ideas. This finding, however, requires further research, and our observations in design studios suggest that these visual stimuli are effective in the creation of inspirational resources and exploration of the unknown in the early stages of the design process.

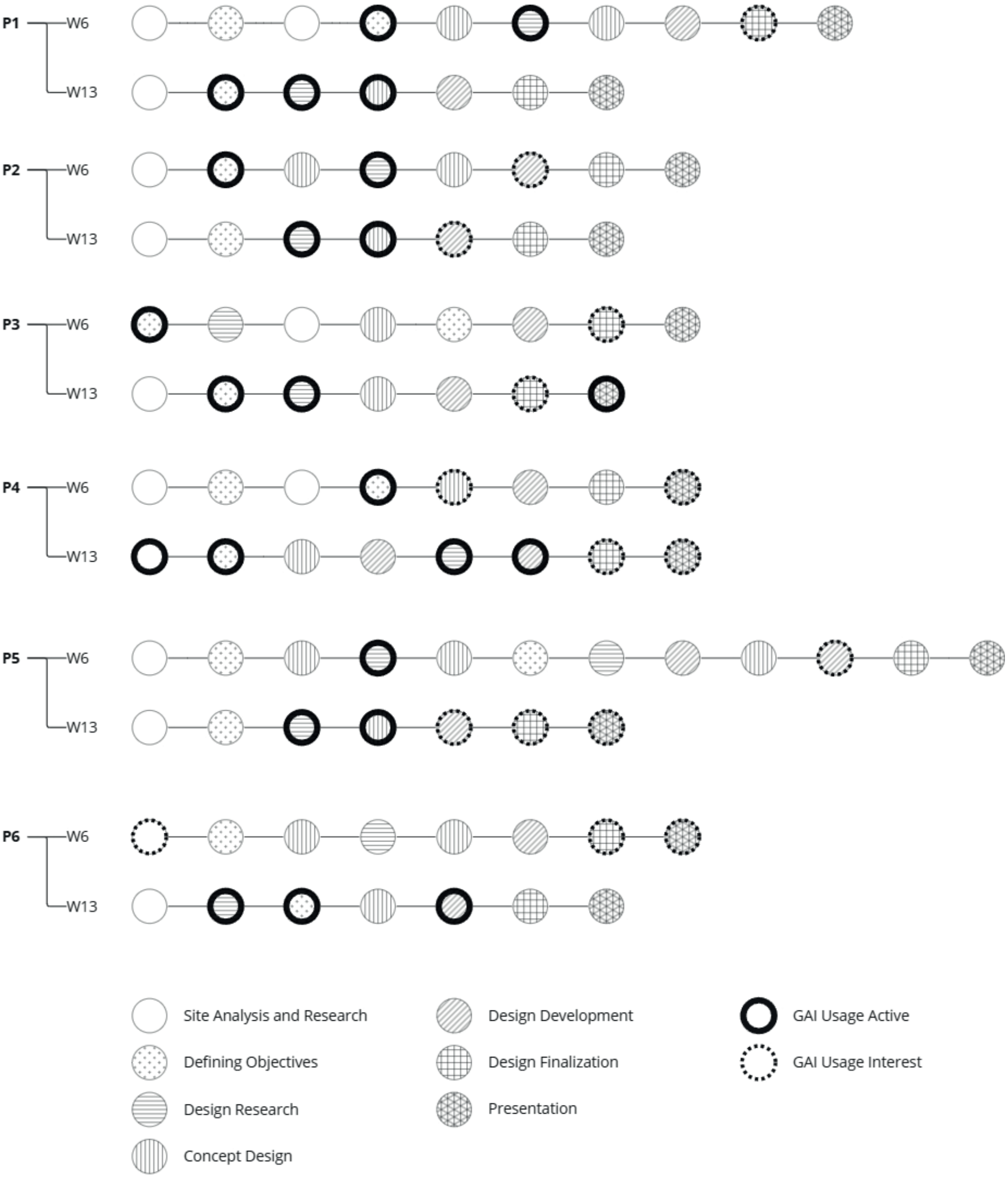

***Figure 7.*** *Coded design process from the semi-structured interviews with the focus group (W6) in sixth week, (W13) in thirteenth week (P.participant).*

A second approach, analogous to Boden's 'combinatorial' method, which entails the generation of new combinations of established ideas, has been shown to be effective as 'solution space enhancement'. This enhancement is evident in the process of generating alternatives that encompass the solution space during the design and development phase. Evidence suggests that students utilize GenAI to advance their preliminary solution ideas. In order to circumvent the potential pitfalls of design fixation (Jansson & Smith, 1991), it is imperative to possess the competencies and the discernment to differentiate between GenAI models and utilize them effectively. On the other hand, Wadinambiarachchi et al. (2024) highlighted that exposure to GenAI images can narrow designers' focus by inadvertently tying them to specific aesthetic elements. This guidance can prevent the exploration of alternative possibilities and constraints the creative process. Students who demonstrated proficiency in seamlessly transitioning between diverse tools and reintegrating salient components into the design cycle exhibited enhanced efficacy. However, some students relinquished the utilization of tools when they encountered difficulties in articulating and completing the process, owing to their inability to effectively visualize the absent components of their design. In this context, conducting a comparable study involving expert designers, while varying the constraint and solution space in different contexts, could serve as a potential avenue for future research.

Another significant application is the process of 'framing and rephrasing', which elevates LLMs and prompts from mere tools to central figures in the design process. Tsidylo and Sendras' (2023) research indicates that Midjourney substantially fosters creativity and innovative thinking in design students by offering comprehensive instructions on image generation prompts. The experience of ideas through text differs significantly from exposure to visual stimuli. Vygotsky (1989) explained that adults can grasp words as concepts, inferring their essence or broader meaning. Iser (1978) suggests that through the act of interpretation, readers construct stories in response to the texts they read, in essence significantly transforming the original text. At the earliest stage, design ideas can be expressed verbally, but these ideas cannot be immediately communicated as a visual image. Again, as suggested in previous studies (Purcell et al., 1993; Perttula, 2006), early exposure to visual images can sometimes restrict one's search for new images (to the point of fixation), at which point encouragement in LLM and GenAI may leave a wider scope for manipulation in the process of transforming a design idea into visual images. Yet, observations suggest that some students find it more beneficial to articulate their ideas verbally. One noteworthy instance involved a student who adopted a comprehensive approach, basing her entire design methodology on the transformation of GenAI visuals into virtual spaces. She employed LLMs to semantically describe these spaces, generating two-dimensional images and subsequently converting them into 3D models. These models were then used to produce plans and sections through the application of appropriate combinations. While this approach necessitates a substantial relinquishment of control, it establishes a novel focal point for the exploration of creativity within the design process. It is acknowledged that further research is necessary to substantiate this claim.

The final approach, which may be considered the most common, involves the utilization of GenAI models as a computer-aided design environment for 'a tool for design externalization' subsequent to the formulation of design decisions. This method, analogous to a rendering engine, demonstrates the efficacy of employing these tools during the documentation stage in achieving substantial time savings. A notable limitation identified at this stage is the propensity to generate images that exhibit a comparable stylistic tendency. The prevalence of commonly and widely used images poses a significant challenge for designers seeking to cultivate an original and creatively distinctive architectural language. Within the context of this challenge, collaboration with GenAI–whether to integrate the designer's distinct style or to utilize a repository of their individual styles–may offer a more advantageous alternative in fostering creative output.

A central limitation shared by these four approaches to leveraging GenAI is the inherently subjective nature of creativity, which resists objective evaluation. The mental connections formed in discovering something new may not be universally recognized as creative. In this exploratory process, when the final output diverges from our initial expectations—whether in form, content, or outcome—we are more inclined to regard it as creative. Conversely, when it merely aligns with what we already anticipate, it tends not to evoke an impression of originality and is therefore less likely to be deemed creative. As a result, creativity is context-dependent: what may strike one observer as innovative could appear banal to another. For instance, a design student's work might seem inventive to peers with limited exposure to examples in that domain, while an experienced design professor, drawing on a broader repertoire, might view the same output as unremarkable.

We posit that this study, coupled with other inquiries into the role of AI in design creativity, holds utility for designers, particularly within the realm of design education. In our studio, we impart various methods and the utilization of AI models as design aids, acknowledging that instructing students in the generation of best creative ideas remains a challenge. Introducing advanced design aids like GenAI models and fostering an awareness of their value in the design exploration process can contribute to the development of work habits that enhance the quest for design alternatives and concepts. This approach may also inspire novices to initiate the construction of their own repository of inspirational resources, potentially beneficial in advancing creative design ideas. Expert designers already curate their pool of designs, occasionally advising novices to craft inspiring collages as a foundation for further development. Extending these endeavors to encompass the use of AI in architectural design holds the promise of enrichment and productivity.

**Acknowledgment**
This research is supported by the ITU Scientific Research Project Unit (İTÜ BAP), designated MGA-2023-44787 and entitled "Enhancing creativity through AI collaboration: What is the role of Generative AI in architectural design education?". Necessary approvals from the ITU Social and Scientific Research and Publication Ethics Board and consents from participating students were obtained (Approval number: 418). We would like to express our gratitude to İTÜ BAP and the students who participated in the studios.